\documentclass[11pt]{article}
\usepackage[a4paper]{geometry}
\usepackage[parfill]{parskip} 
\usepackage{graphicx}
\usepackage{amsmath}
\usepackage{algorithmic}
\usepackage{algorithm}
\usepackage{subfigure}
\usepackage{tikz}
\usepackage{cite}

\def\degree{{k}}
\def\linksRC{k^+}

\def\degree{k}
\def\eProb{p}
\def\entropy{S}
\def\stepFunct{w}

\def\walk{W}

\begin{document}
\begin{center}
{\LARGE
Network partition via a bound of the spectral radius \\[1ex]}
{R. J. Mondrag\'on\\[2ex]
Queen Mary University of London\\
School of Electronic Engineering and Computer Science\\
Mile End Road\\
E1 4NS, London\\
21 May 2014\\ 
}
\end{center}

\section*{Abstract}
Based on the density of connections between the nodes of high degree, we introduce two bounds of the spectral radius. We use these bounds to split a network into two sets, one of these sets contains the high degree nodes, we refer to this set as the spectral--core.  The degree of the nodes of the subnetwork formed by the spectral--core gives an approximation to the top entries of the leading eigenvector of the whole network. We also present some numerical examples showing the dependancy of the spectral--core with the assortativity coefficient, its evaluation in several real networks and how the properties of the spectral--core can be used to reduce the spectral radius.
\section{Introduction}
In some networks its partition into two substructures occurs naturally as each partition plays a distinctive role. The best known of these partitions is the core--periphery introduced by Borgatti and Everett~\cite{Borgatti2000models}, where the core is a set of nodes that are densely interconnected and share connections with the periphery nodes. In contrast the periphery nodes are poorly interconnected. The idea behind this partition is that the core nodes dominate the behaviour of the whole network~\cite{Borgatti2000models}. There exist many practical methods to evaluate the core--periphery~\cite{Borgatti2000models,holme2005core,cucuringu2014detection,avin2014core,lee2014density,barucca2015centrality,zhang2015}.

Another way to partition a network is to use the rich--club~\cite{zhou04}. The rich--club is the set of well connected nodes that are well interconnected, and they form a rich--core~\cite{ma2015rich}. 
The rich--core is based on the idea that the well connected nodes tend to dominate network properties like  assortativity and clustering coefficient~\cite{Xu2010}. Compared to the core--periphery, the rich--core definition does not impose any restriction on the poorly connected nodes, the periphery.

In here, we introduce a partition using a bound of the spectral radius. The connectivity of an undirected and unweighted network can be represented with the adjacency matrix $\bf A$ where $A_{ij}=A_{ji}=1$ if nodes $i$ and $j$ share a link and $A_{ij}=0$ otherwise. 
The spectrum of the adjacency matrix is the set of eigenvalues $\Lambda_1 \ge \Lambda_2\ge \ldots \ge \Lambda_N$ where $\Lambda_1$ is the spectral radius.  This first eigenvalue $\Lambda_1$ plays an important role when describing information diffusion or epidemic transmission on a  network~\cite{wang2003epidemic,gomez2010discrete,youssef2011individual,van2011n}.
 For example, in the SIS epidemic model, the epidemic threshold where the steady-state of infected nodes changes from no--infected to all infected is determined by the inverse of largest eigenvalue $1/\Lambda_1$. Based on the idea of the rich--club we introduce a bound for $\Lambda_1$ which is related to the density of connections between the best connected nodes. The maximal value of this density defines the core. 

Next section introduces a new spectral bound and the definition of the spectral--core. In Section~3 we present some examples  to show; the dependance of the bound with the assortativity coefficient, when the network cannot be partitioned and, how the properties  of the core give a good approximation to the leading eigenvector. The final section is our conclusions.

\section{Network partition via a spectral bound}
In a network where the nodes are ranked in decreasing order of their degrees, the connectivity of the network can be represented with the degree sequence $\{\degree_i\}$ and the sequence of number of links $\{ \linksRC_r \}$ that node $r$ shares with nodes of higher rank.  This last sequence, $\{\linksRC_i\}$ is bounded by the degree $\linksRC_i \le \degree_i$ and satisfies that $\sum_{r=1}^N\linksRC_r=L$, where $N$ is the total number of nodes and $L$ is the total number of links. 
Here we propose a bound for $\Lambda_1$  from the $\{ \linksRC_i \}$ sequence. A lower bound for $\Lambda_1$ is~\cite{van2010influence,van2010graph}
$\Lambda_1 \ge \left( \walk_{2n}/\walk_0\right)^{1/(2n)}\ge  \left( \walk_{n}/\walk_0\right)^{1/n}, \quad n=1,\ldots$
where $\walk_n= \overline{u}^T{\bf A}^n \overline{u}$ is the total number of walks of length $n$, ${\bf A}$ is the adjacency matrix and $\overline{u}$ is a vector with all its entries equal to one. An upper bound for the number of walks is $\walk_n\le \sum_{j=1}^{N}\degree_j^n$~\cite{fiol2009number} where the equality is true only if $n\le 2$. Using the bound for $\walk_n$ for $n=1$ we define
\begin{eqnarray}
\nonumber
g(r)&=& \frac{1}{r}\sum_{i=1}^r \degree_i
= \frac{1}{j}\sum_{i=1}^r \left( 2\linksRC_i + \degree_i -2\linksRC_i\right)\\
&=& \frac{1}{r}\sum_{i=1}^r  2\linksRC_i  + \frac{1}{r}\sum_{i=1}^r \left( \degree_i -2\linksRC_i\right)\\
\nonumber
&=&2 \langle \linksRC\rangle_r +  \langle \degree-2\linksRC\rangle_r
,\quad r=1,\ldots , N
\end{eqnarray}
which is the average number of links for the top $r$ ranked nodes. The sum containing only terms of the form $2\linksRC_i$ gives the average number of links between the top $i$ ranked nodes. The other sum containing the terms $\degree_i-2\linksRC_i$ is the average number of links between the top $i$ ranked nodes and nodes of lower rank. Notice that  if $r=N$ then $2 \langle \linksRC\rangle_N =2L/N$, $\langle \degree-2\linksRC\rangle_N=0$ and $g(N)=2L/N$,  which is the well known lower bound of $\walk_1/\walk_0 = 2L/N\le \Lambda_1$.
Also notice that $2 \langle \linksRC\rangle_r$  could be larger than $g(N)=2L/N$. 
We split the network into two parts by considering  the value $r$ such that $2 \langle \linksRC\rangle_r$ is maximal, that is when the density of connections between the top ranked nodes is maximal. In this case the core of the network is the nodes of rank greater than $r_c$  where
\begin{equation}
r_c = \underset{r}{\mathrm{argmax}} \left(2 \langle \linksRC\rangle_r\right)
\end{equation}
and the bound is
\begin{equation}
\Lambda_1 \ge b_1 \equiv 2\langle \linksRC \rangle_{r_c}.
\label{eq:linearBound}
\end{equation}
The core, that we refer here as the spectral--core, are the nodes $n_i$ in the subset $\{n_1,\ldots ,n_{r_c} \}$.
To confirm that $2\langle \linksRC \rangle_{r_c}$ is a bound of the spectral radius consider Rayleigh's inequality $\Lambda_1\ge \overline{u}^T{\bf A}\overline{u}/(\overline{u}^T\overline{u})$. If ${\bf A}$ is the adjacency matrix of the network ranked in decreasing order of its node's degree and  $\overline{u}$ is a vector with ones in the top $r_c$ entries and zero otherwise then Rayleigh's inequality gives $\Lambda_1 \ge 2\langle \linksRC \rangle_{r_c}$.
Also notice that the components of  $\overline{y}={\bf A}\overline{u}$ are $y_i=\degree^{({r_c})}_i$ which are the degree of the nodes of a network that consist of only the top $r_c$ ranked nodes and $\overline{y}$ is an approximation of the top $r_c$ entries of the eigenvector $\overline{v}_1$, where  ${\bf A}\overline{v}_1=\Lambda_1 \overline{v}_1$. This approximation is good if the spectral gap $|\Lambda_1-\Lambda_2|$ is large.

The previous bound can be improved if we consider the case $k=2$. 
Following the same procedure as for the case $k=1$ we obtain that the partition is defined by
\begin{equation}
r_c = \underset{r}{\mathrm{argmax}} \left(\langle (2\linksRC)^2\rangle_r\right).
\end{equation}
The bound  is $\Lambda^2_1\ge \overline{u}^T{\bf A}^2\overline{u}/(\overline{u}^T\overline{u})$ where $\overline{u}$ is a vector with the top $r_c$ entries are ones and the rest are  zeros gives
\begin{equation}
\Lambda_1 \ge b_2 \equiv \left( \frac{\sum_{i=1}^{{r_c}} (2\linksRC_i)^2}{r_c}\right)^{1/2}=2\langle (\linksRC)^2\rangle_{r_c}^{1/2}.
\label{eq:quadraticBound}
\end{equation}

Notice that it is possible that the value of $r_c$,  for both bounds, be equal to the total number of nodes $N$. This would happen for a network where the density of links of the $r$ top ranked nodes is always lower than the overall density of links in the network, that is $\sum_{j=1}^r\linksRC_i < r(2L/N)=r\langle k \rangle_N$ for $r=1,\ldots ,N$. These networks would not be partitioned by our method or in other words the spectral--core would be the whole network.


\section{Examples}
\begin{figure}
\begin{center}
\subfigure[]{
\includegraphics[width=6cm]{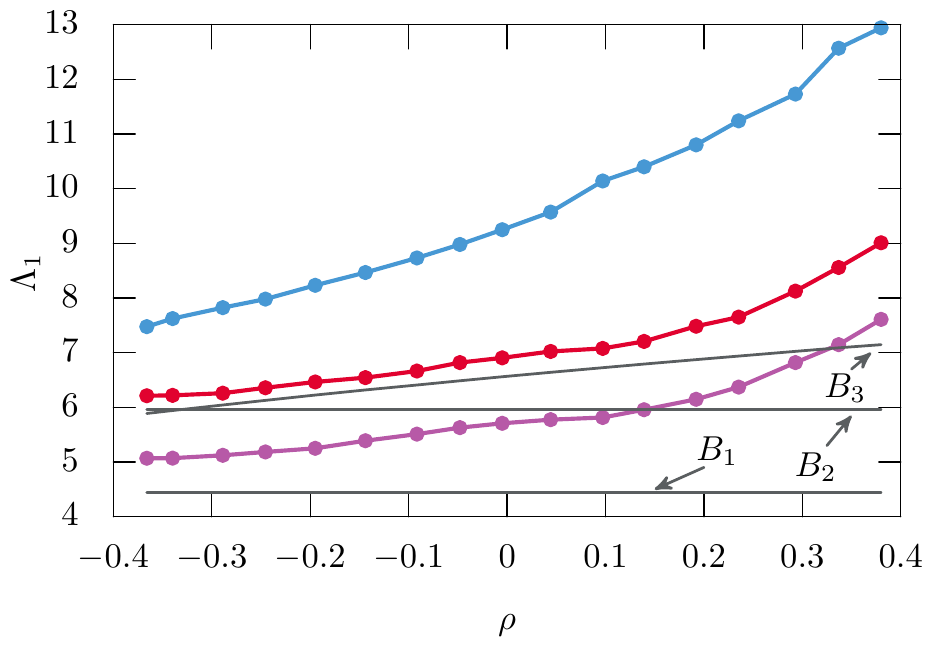}
}
\subfigure[]{
\includegraphics[width=6cm]{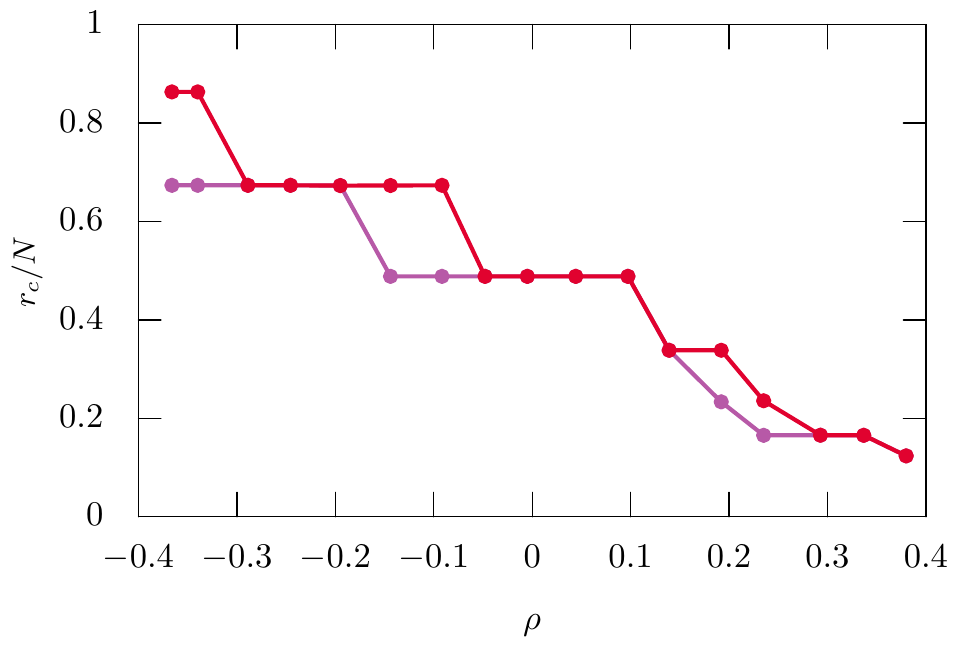}}\\
\subfigure[]{
\includegraphics[width=6cm]{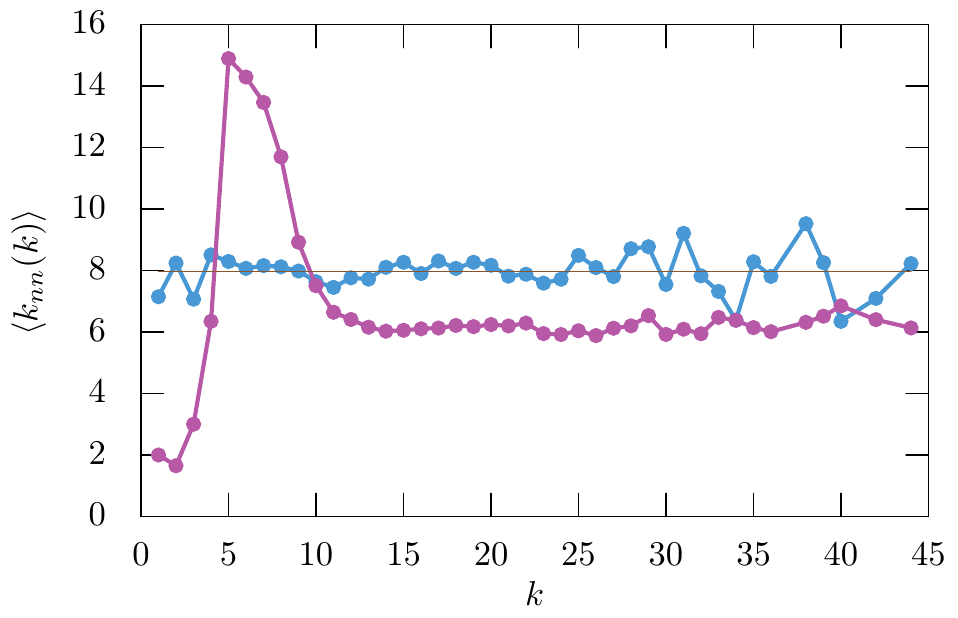}
}
\subfigure[]{
\includegraphics[width=6cm]{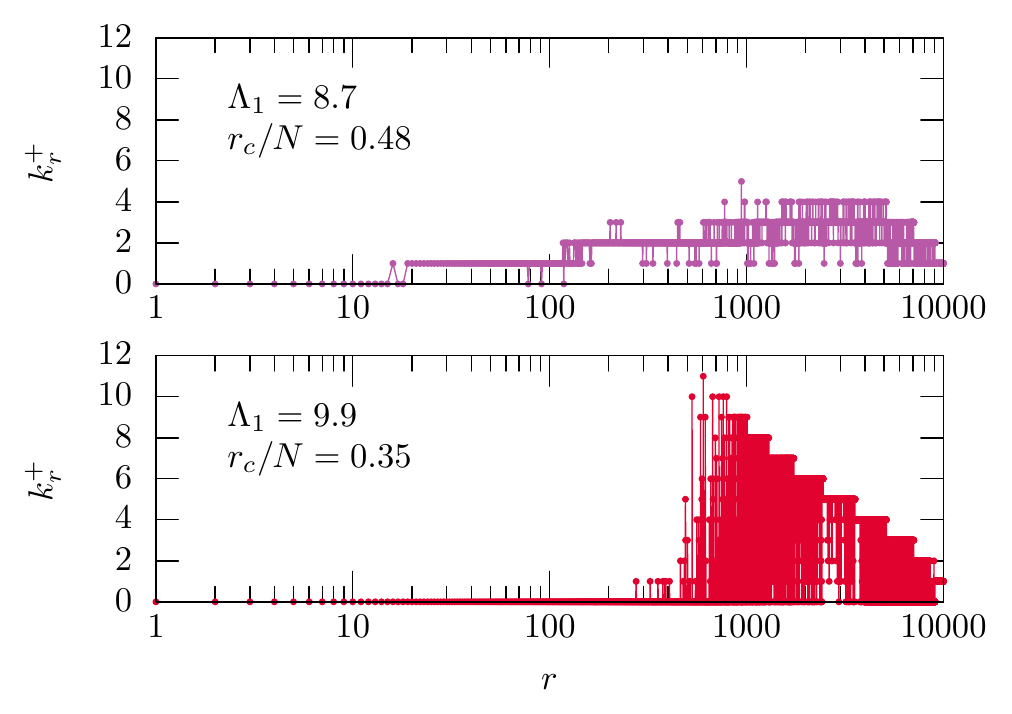}}
\end{center}
\caption{\label{fig:spectraEi} (a) Spectral bound $\Lambda_1$ (blue) and its bounds $b_1$ (Eq.~(\ref{eq:linearBound})) purple and $b_2$ (Eq.~\ref{eq:quadraticBound}) (red) as a function of the assortativity coefficient $\rho$. (b) Relative size of the spectral--core for the bound $b_1$(purple) and $b_2$ (red). (c) Average neighbours degree $k_{nn}(k)$ for two networks with $\rho=0$, the straight line is the value of $k_{nn}(k)=\langle k^2 \rangle/\langle k_r \rangle$ for the uncorrelated network. (d) The sequence $\{\linksRC\}$ for the networks with equal assortativity but different spectral properties. The networks in (c) and (d) were  obtained using the maximal entropy approach, where the degree sequence $\{ \degree_r\}$ is fixed, see Appendix.
}
\end{figure}

It is known that increasing the assortativity of a network also increases the spectral radius and hence its lower bound~\cite{van2010influence,d2012robustness} which is the case for the  bound presented here.
Fig~\ref{fig:spectraEi}(a) shows the spectral radius and the bounds for a generated network with a target assortative coefficient. The network has a power law degree distribution $P(k)\sim k^{-2.8}$ with $N=10\,515$ and $L=23\,375$. For comparison the figure also shows the bounds $B_1=\walk_1/\walk_0=2L/N$, $B_2=(N\walk_2/\walk_0)^{1/2}$ and $B_3=(\walk_3/\walk_0)^{1/3}$.
Fig.~\ref{fig:spectraEi}(b) shows that the relative size of the core $r_c/N$ is large if the network is disassortative and decreases as the network becomes more assortative.  As expected the bound $b_2$ is better than $b_1$ but sometimes at the expense that $b_2$ requires a larger spectral--core.
Fig.~\ref{fig:spectraEi}(a) suggests that positive assortativity means small spectral--core. However, we have to be careful with this assertion as there is ambiguity when classifying networks using the assortativity coefficient. Different networks can have the same assortative coefficient even if they are very differently connected~\cite{dorogovtsev2010lectures}. Fig.~\ref{fig:spectraEi}(c) shows the average neighbours degree $k_{nn}(k)$ for two networks with identical degree sequence and assortativity coefficient $\rho=0.005$ but with different  connectivity as shown by their $\{\linksRC_i\}$ sequence (Fig.~\ref{fig:spectraEi}(d)). 

Figure~2 (a)--(b) shows the spectral radius and its bounds and the relative spectral--core size of some real networks.
Notice that the relative core size is not an indicator if the network is assortative or not. The Hep-Th network is assortative and the AS--Internet is disassortative, and both networks have very small core when considering the bound $b_1$. The Football network is an example of a network where the bounds $b_1$ and $b_2$ are the same as the bounds obtained when considering all the network nodes, that is $r_c=N$. The reason is that the average number of links of the $r$ ranked nodes for all $r$ is smaller than $2L/N$.
 
\begin{figure}
\begin{center}
\subfigure[]{\includegraphics[width=6cm]{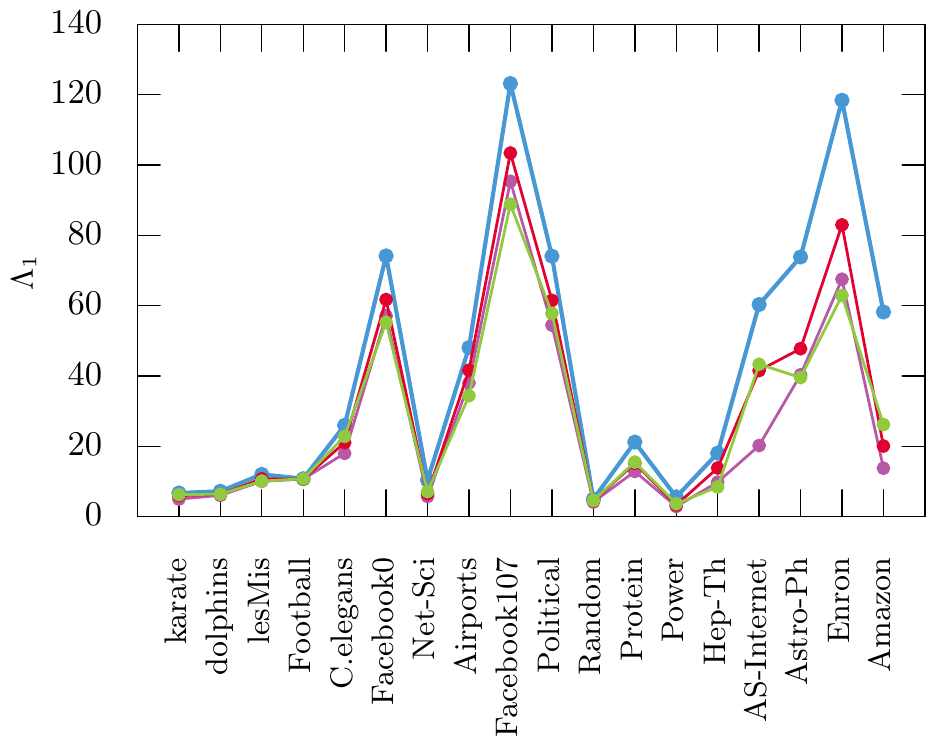}}
\subfigure[]{\includegraphics[width=6cm]{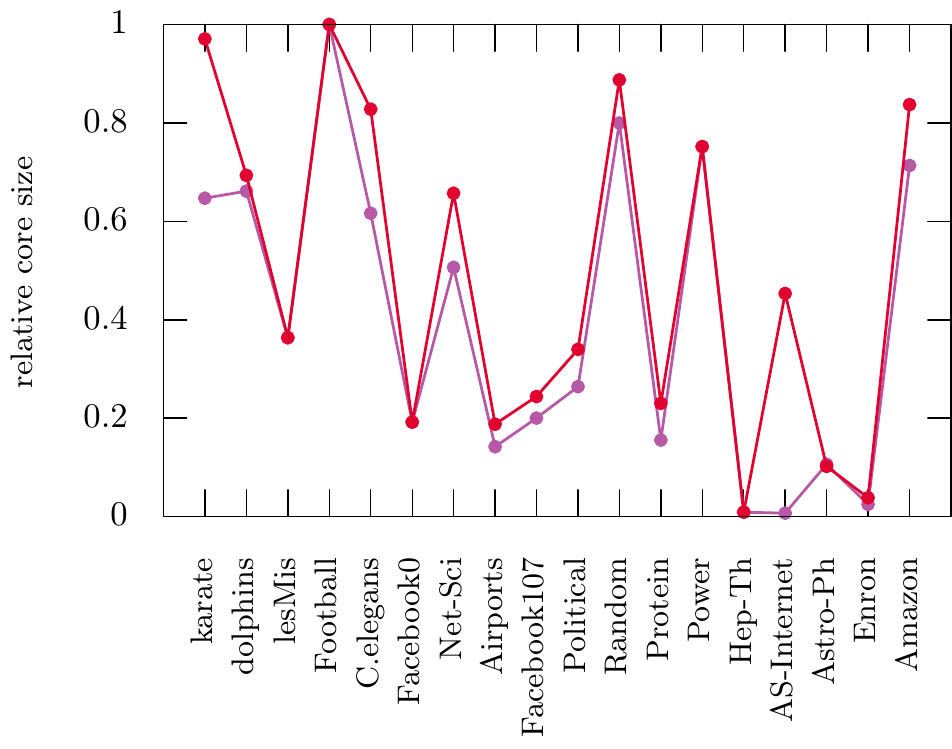}}
\end{center}
\caption{\label{fig:realNets} (a) The spectral radius (blue) and its bounds for several real networks. The linear bound $b_1$ (purple), $b_2$ (red) and $(N_3/N_0)^{1/3}$. (b) The relative size of the core obtained from the bounds $b_1$ (purple) and $b_2$ (red).}
\end{figure}

\begin{figure}
\begin{center}
\subfigure[]{\includegraphics[width=6cm]{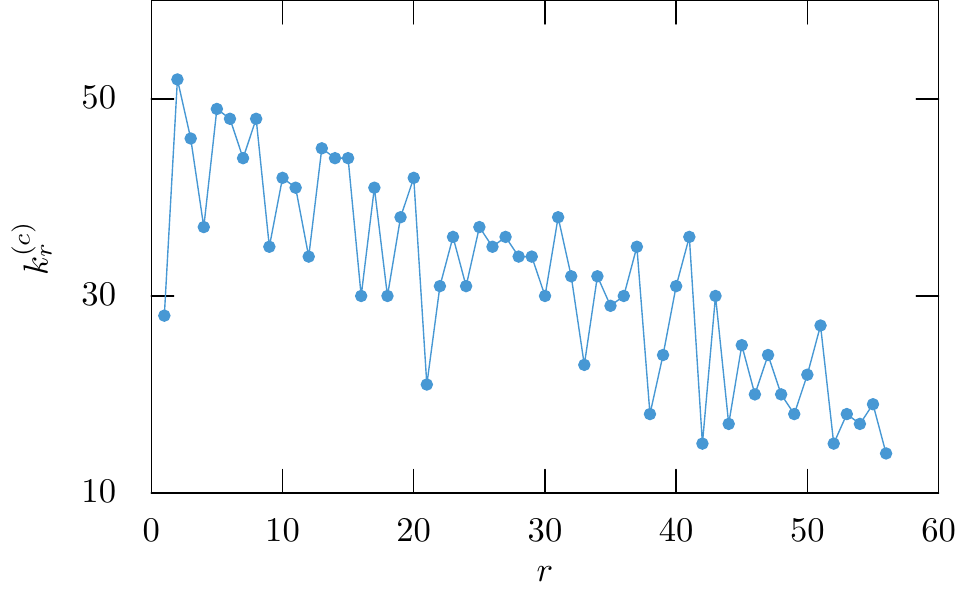}}
\subfigure[]{{\includegraphics[width=6cm]{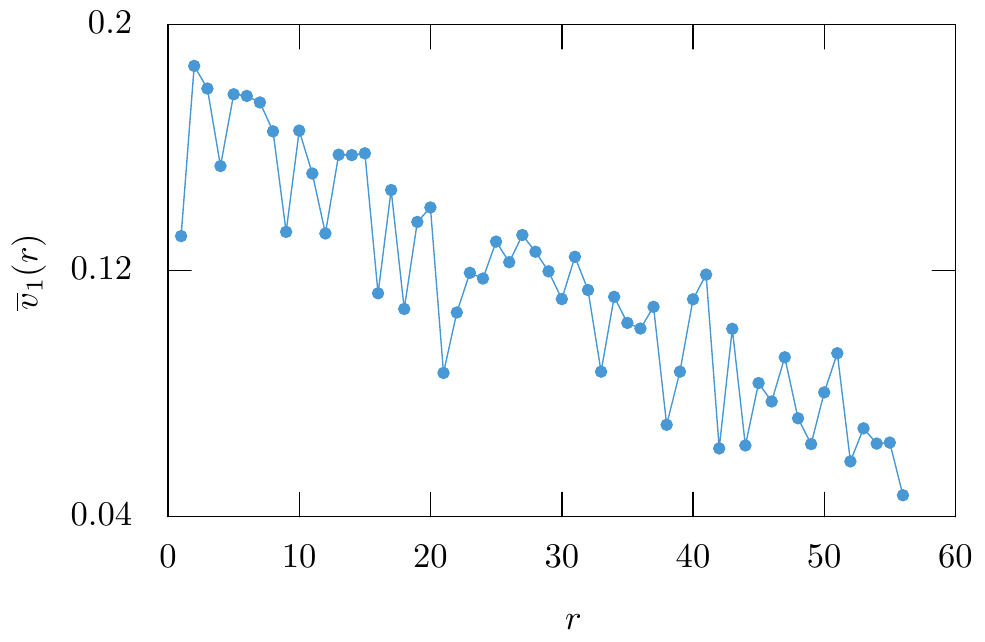}}}\\
\subfigure[]{{\includegraphics[width=6cm]{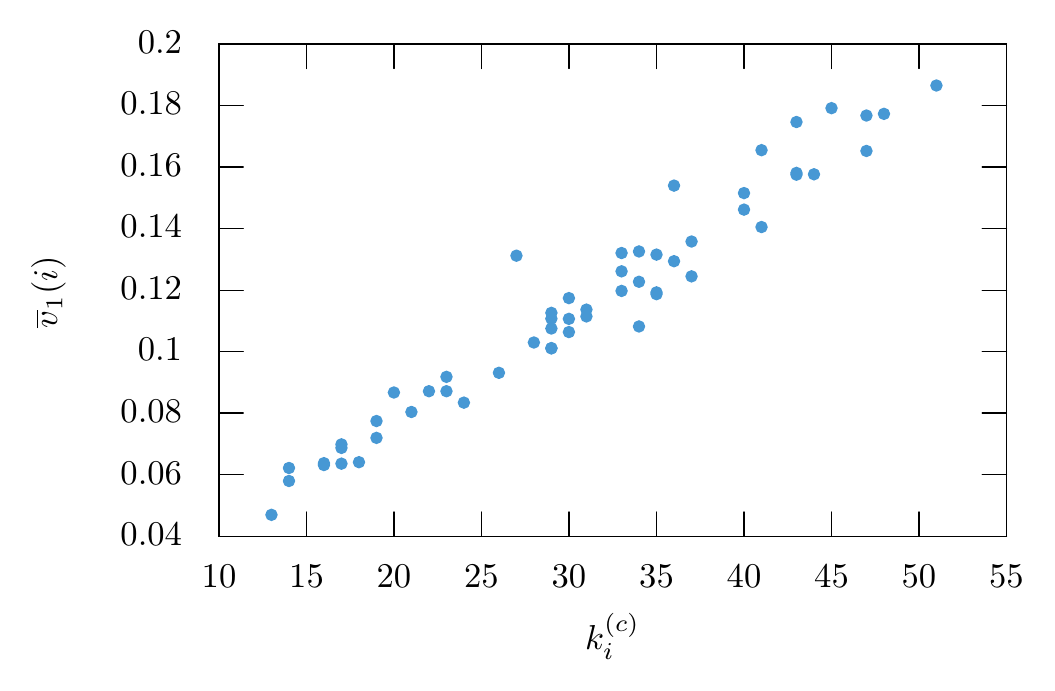}}}
\subfigure[]{{\includegraphics[width=6cm]{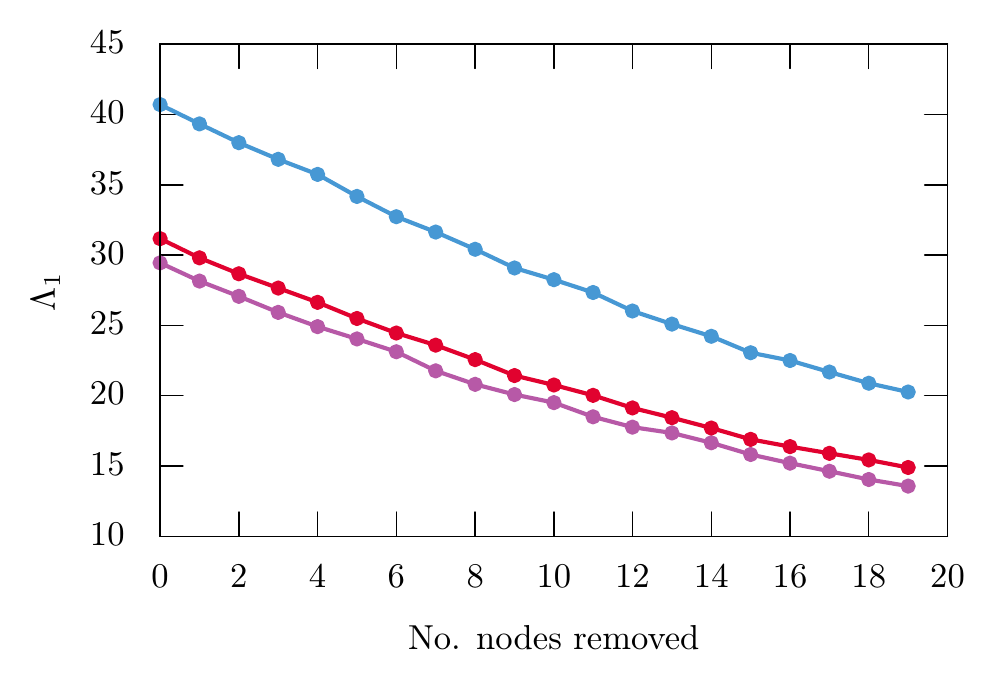}}}
\end{center}
\caption{\label{fig:removeNodes} (a) Degree of the subnetwork formed by the nodes in the core and (b) first $r_c$ entries of the leading eigenvector for the EU--airports network. (c) Correlation between $k_r^{(c)}$ and $\overline{v}_1(r)$. The correlation coefficient between these two quantities is $0.97$. (d) Change of the spectral radius and its bounds as the high degree nodes in the spectral--core are removed. }
\end{figure}

Our final example is the european airports network. The network is formed by the aggregation of 34 airlines and their destinations within Europe. The network has $N=417$ nodes, which is the number of different airports,  and $L=2\,953$ links which is the number of unique routes between airports~\cite{cardillo2013emergence}. 
A common question when considering a transport network is, if a virus was spreading via the airports network, which airports should stop operating to reduce the chances of an epidemic, in other words which nodes should be remove to decrease $\Lambda_1$.  A common procedure is to evaluate the leading eigenvector and remove the nodes that correspond to the highest entries of this eigenvector. Figure~\ref{fig:removeNodes}(a) shows the degree $\degree^{(c)}_r$ of the network formed only by the nodes contained on the spectral--core defined by the bound $b_1$. This network contains only 57 nodes. Figure~\ref{fig:removeNodes}(b) shows the first 57 top entries of the leading eigenvector when considering the whole network. It is clear that the degree of the network formed by the core and these eigenvector entries are correlated (Fig.~\ref{fig:removeNodes}(c)). 
Hence the strategy for reducing $\Lambda_1$ using the leading eigenvector is very similar to the strategy of removing the nodes with the highest degree on the spectral--core network. 
Fig.~\ref{fig:removeNodes}(d) shows how $\Lambda_1$ decreases as the highest degree of the spectral--core network are removed, which gives very  similar results to the procedure of removing the highest entries of  the leading eigenvector (not shown in the Figure). 

\section{Conclusion}
Networks that have well interconnected hubs can be partitioned into two parts. The partition is done by considering a bound of the spectral radius, the bound is based on the density of connection between the hubs. We refer to the subset containing the hubs as the spectral--core. The degree of the nodes of the subnetwork containing the spectral--core gives an approximation to the entries of the leading eigenvector of the whole network. For networks where the hubs are well interconnected, the bound obtained by considering only the hubs is tighter  than its equivalent bound when considering the whole network.
Notice that the sequence $\{\linksRC_i \}$ seems to play an important role when describing undirected networks as it has been used to evaluate, in closed form, the maximal entropy of an undirected, unweighted network~\cite{mondragon2014a}, the definition of the rich--core~\cite{ma2015rich} and now spectral properties.

\bibliographystyle{unsrt}

\newpage
\section{Appendix}
The maximisation of the network entropy can be used to create a network ensemble that is Ômaximally nonÐcommittalÕ, that is given the $\{\degree_1,\ldots ,\degree_N\}$ and $\{\linksRC_1, \ldots, \linksRC_N\}$ constraints, the ensemble is as unbiased as it is possible.
%
The Shannon entropy of the network is $\entropy = -\sum_{i=1}^N \sum_{j=1; j\ne i}^N \eProb_{ij} \log( \eProb_{ij} )$. The maximal entropy solution is given by the probabilities~\cite{mondragon2014a}
\begin{equation}
\label{eq:mainResult}
\eProb_{ij} = 
\frac{\stepFunct(i)\left(\degree_i-\linksRC_i \right) }{\sum_{n=1}^{j-1} \stepFunct(n)\left(\degree_n-\linksRC_n \right)}\frac{ \linksRC_j}{L}\quad i<j\\
\end{equation}
where 
\begin{equation}
\label{eq:stepFunct}
\stepFunct(m) = \frac{\stepFunct(m-1)\sum_{i=1}^{m-1}\stepFunct(i)(\degree_i-\linksRC_i)}{\sum_{i=1}^{m-1}\stepFunct(i)(\degree_i-\linksRC_i)-\linksRC_m\stepFunct(m-1)}.
\end{equation}
The values of $\stepFunct(m)$  are defined recursively with the initial condition $\stepFunct(1)=1$. The average number of links between nodes $i$ and $j$ is $e_{ij}=Lp_{ij}$ with variance $s_{ij}=Lp_{ij}(1-p_{ij})$.

To evaluate the network properties from the ensemble we use $w(m)$ and $G(m)= \sum_{i=1}^{m-1}w(i)(\degree_i-\linksRC_i)$.

\begin{algorithm}
\caption{Sequences:$w(m)$ \& $G(m)$}
\begin{algorithmic}[1]
\REQUIRE $\degree_m$ and $\linksRC_m$ for $m=1,\ldots, N$
\STATE $\stepFunct(1)=1$
\STATE $G(1) \leftarrow w(1)(\degree_1-\linksRC_1)$
\FOR{$m=2$ to $N$}
\STATE $\stepFunct(m) \leftarrow \stepFunct(m-1)*G(m-1)/(G(m-1)-\linksRC_m*\stepFunct(m))$
\STATE $G(m) \leftarrow G(m-1)+\stepFunct(m)*(\degree_m-\linksRC_m)$
\ENDFOR
\RETURN $\{w\}$ and $\{G\}$
\end{algorithmic}
\end{algorithm}

The assortativity is evaluated using~\cite{Newman02}
\begin{equation}
\rho =\frac{\left\langle k k' \right\rangle_\ell- \left\langle k \right\rangle^2_\ell}{\left\langle k^2\right\rangle_\ell-\left\langle k  \right\rangle_\ell^2}
\end{equation}
with
\begin{equation}
\left\langle k \right\rangle_\ell=\sum_i\sum_{j\ne i} \degree_i p_{ij}=\frac{\langle k^2 \rangle_n }{ \langle k \rangle_n }
\end{equation}
where $\langle \ldots\rangle_\ell$ is the average over all links and $\langle \ldots \rangle_n$ is the average over all nodes.
The average degree of the end nodes of a link is $\left\langle k k' \right\rangle_\ell = \sum_i\sum_{j\ne i} \degree_i \degree_j p_{ij}$. 
Then 
\begin{equation}
\rho = 2\frac{T_1-T_2T_2}{T_3-T_2T_2}
\end{equation}
where $T_0=p_{ij}=w(j)(\degree_j-\linksRC_j)\degree_i/G_j$, $T_1=\langle k k'\rangle=\sum_{i=1}^N\sum_{j=1}^N p_{ij}k k'$, $T_2=\langle k\rangle = \sum_{i=1}^N\sum_{j=1}^N p_{ij}(\degree_i+\degree_j)/2$ and $T_3=\langle k^2 \rangle=\sum_{i=1}^N\sum_{j=1}^N p_{ij}(\degree_j^2+\degree_i^2)/2$. The algorithm is named Assortativity~coefficient. Notice that in the algorithm the inner sum is evaluated from $j=1$ to $i-1$ and there is a factor of 2, this is because we are using the fact that the network is undirected, i.e. $p_{ij}=p_{ji}$.
\begin{algorithm}
\caption{Assortativity coefficient from the ensemble}
\begin{algorithmic}[1]
\REQUIRE $\degree_m$ and $\linksRC_m$ for $m=1,\ldots, N$
\STATE evaluate $w(m)$ and $G(m)$ using algorithm Sequence: $w(m)$ \& $G(m)$
\STATE $T_0 \leftarrow 0$,$T_1 \leftarrow 0$,$T_2 \leftarrow 0$,$T_3 \leftarrow 0$ 
\FOR{$i=1$ to $N$ }
\FOR{$j=1$ to $i-1$ }
\STATE $T_0 \leftarrow \stepFunct_j (\degree_j-\linksRC_j)\linksRC_i/G_j$
\STATE $T_1 \leftarrow T_1+T_0\degree_i\degree_j$
\STATE $T_2 \leftarrow T_2+ T_0(\degree_i+\degree_j)/2$
\STATE $T_3 \leftarrow T_3+T_0(\degree_j\degree_j + \degree_i\degree_i)/2$
\ENDFOR
\ENDFOR
\RETURN $\rho \leftarrow 2(T_1-T_2T_2)/(T_3-T_2T_2)$
\end{algorithmic}
\end{algorithm}

To evaluate an ensemble with a given assortativity coefficient we use Simmulated Annealing. The input is the sequences $\{\degree_i\}$ and $\{\linksRC_i\}$ from the original network, the output is a new sequence $\{\linksRC_i \}_{(new)}$ such that the ensemble obtained from $\{\degree_i\}$ and  $\{\linksRC_i \}_{(new)}$ has the target assortativity coefficient. Notice that the degree sequence is always conserved.
\begin{algorithm}
\caption{Ensemble target assortativity}
\begin{algorithmic}[1]
\REQUIRE $\degree_m$  and $\linksRC_m$ for $m=1,\ldots, N$. 
\REQUIRE target assortativity coefficient $\rho_T$
\REQUIRE Precision of the solution $\epsilon=0.001$, $T$ is the temperature parameter in the Monte-Carlo step
\STATE $\Delta E \leftarrow 1$
\STATE $T=10$
\STATE $iCount=1$\COMMENT{counts successful step in the Monte Carlo step}
\WHILE{$\Delta E > \epsilon$} 
\STATE Select nodes $i$ and $j$ at random, where $i\ne j$
\IF{ $\linksRC_i<\degree_i$ and $\linksRC_i<i-1$ and $\linksRC_j\ge 1$} 
\STATE $\linksRC_i \leftarrow \linksRC_i +1$
\STATE $\linksRC_j \leftarrow \linksRC_j -1$
\STATE $\rho_{new} \leftarrow \rho$, evaluate the new assortativity using algorithm Assortativity
\STATE $\Delta E \leftarrow |(\rho_{new}-\rho_T)|$
\COMMENT {Monte--Carlo step}
\IF{$\Delta E > 0 $ or $\Delta E > \exp(-random()/T)$ } 
\STATE $\linksRC_i \leftarrow \linksRC_i -1$
\STATE $\linksRC_j \leftarrow \linksRC_j +1$
\ELSE
\STATE $iCount \leftarrow iCount+1$
\ENDIF
\ENDIF
\COMMENT{Reduce temperature every 1000 successful steps}
\IF {$iCount \% 1000$}
\STATE $T \leftarrow T*0.85$
\STATE $iCount \leftarrow 1$
\ENDIF
\ENDWHILE
\RETURN $\{\linksRC_i\}_{(new)} \leftarrow \{\linksRC_i\}$
\end{algorithmic}
\end{algorithm}

To construct a network from the ensemble, using the sequences $\{\degree_i \}$ and $\{\linksRC_i \}$ we evaluate the probability $p_{ij}$ (Eq.~(\ref{eq:mainResult} )) that  a link exist between nodes $i$ and $j$. For each node $i$, $\degree_i$ stubs are assigned to the node,  the stubs are divided into the ones that connect to nodes of higher rank, i.e.  $\linksRC_i$ and the rest $\degree_i-\linksRC_i$. Taking two nodes, a stub of the node with lower rank connects with a stub of a node of higher rank. The probability of connection is $p_{ij}$, we do no allow self--loops or multiple links. It is possible that we can end up with a node with two stubs or two nodes that have already a link and each has a free stub. If this is the case the procedure is started again from the beginning until we get a network with all the stubs connected.

\subsection*{Datasets}
The dataset for the networks Karate, dolphins, LesMis, Football, {\sl C. elgans}, Net-Sci (collaboration between Network Scientists) , Political (blogs), Hep-th and Astro-Ph are available from M. Newman's web page ({\verb http://www-personal.umich.edu/~mejn/netdata/}). The datasets amazon (amazon0601), facebook  and  enron (email-Enron) are available from the Stanford Large Network Dataset Collection (https://snap.stanford.edu/data).  The random network is an Erods-Renyi network generated with igraph. The European network is available from Air Transportation Multiplex ({\verb!http://complex.unizar.es/~atnmultiplex/!}).

\end{document}